\documentclass[showpacs,superscriptaddress,english,twocolumn,prb]{revtex4}
\usepackage[T1]{fontenc}
\usepackage[latin1]{inputenc}
\usepackage{graphicx}
\makeatletter

\newcommand{\n}[1]{\ensuremath{\mathrm{#1}}}
\newcommand{\EE}{\end{equation}}
\newcommand{\BE}{\begin{equation}}
\usepackage{babel}
\makeatother
\begin{document}

\title{Vacuum Rabi oscillations in a macroscopic superconducting qubit LC oscillator system}

\author{J. Johansson}
\email{janj@will.brl.ntt.co.jp}
\affiliation{NTT Basic Research Laboratories, NTT Corporation, Atsugi, Kanagawa 243-0198, Japan}
\affiliation{CREST, Japan Science and Technology Agency, Saitama 332-0012, Japan}
\author{S. Saito}
\affiliation{NTT Basic Research Laboratories, NTT Corporation, Atsugi, Kanagawa 243-0198, Japan}
\affiliation{CREST, Japan Science and Technology Agency, Saitama 332-0012, Japan}
\author{T. Meno}
\affiliation{NTT Advanced Technology Corporation, NTT Corporation, Atsugi, Kanagawa 243-0198, Japan}
\author{H. Nakano}
\affiliation{NTT Basic Research Laboratories, NTT Corporation, Atsugi, Kanagawa 243-0198, Japan}
\affiliation{CREST, Japan Science and Technology Agency, Saitama 332-0012, Japan}
\author{M. Ueda}
\affiliation{Department of Physics, Tokyo Institute of Thechnology, Tokyo 152-8551, Japan}
\affiliation{CREST, Japan Science and Technology Agency, Saitama 332-0012, Japan} 
\author{K. Semba}
\affiliation{NTT Basic Research Laboratories, NTT Corporation, Atsugi, Kanagawa 243-0198, Japan}
\affiliation{CREST, Japan Science and Technology Agency, Saitama 332-0012, Japan} 
\author{H. Takayanagi}
\affiliation{NTT Basic Research Laboratories, NTT Corporation, Atsugi, Kanagawa 243-0198, Japan}
\affiliation{CREST, Japan Science and Technology Agency, Saitama 332-0012, Japan} 

\begin{abstract}
We have observed the coherent exchange of a single energy quantum between a flux qubit and a superconducting LC circuit acting as a quantum harmonic oscillator. The exchange of an energy quantum is known as the vacuum Rabi oscillations: the qubit is oscillating between the excited state and the ground state and the oscillator between the vacuum state and the first excited state. We have also obtained evidence of level quantization of the LC circuit by observing the change in the oscillation frequency when the LC circuit was not initially in the vacuum state.
\end{abstract}

\maketitle
The coupling of quantum two--state systems to harmonic oscillators constitutes a broad and active research field, most notably in atomic cavity quantum electrodynamics (QED) where atoms having a large electrical dipole moment interacts with the quantized electromagnetic field in high-Q cavities\cite{cQED}, but also in other systems, for example, ions held in linear traps where the ions couple to a vibrational mode acting as a harmonic oscillator\cite{ions}, and in solid state, where coupling of atom-like objects to harmonic modes has been pursued for some time, starting from quantum wells in micro-cavities\cite{weisbuch} to recent experiments on quantum dots in photonic crystals\cite{Qdot1}, and two experiments with superconducting devices: a system with a flux qubit coupled to a superconducting quantum interference device (SQUID) showed Rabi oscillations driven by an external field\cite{DelftRes} and the vacuum Rabi splitting due to strong coupling of a Cooper pair box to a microwave photon in a superconducting waveguide was demonstrated\cite{YaleRes}. These systems attract increasing attention because of their potential use in future quantum information processing. In addition, they offer an ideal testing ground for studying the fundamental interactions between matter (atoms, qubits) and light (harmonic oscillators). 

\begin{figure}[thb]
  \includegraphics[width=0.95\linewidth]{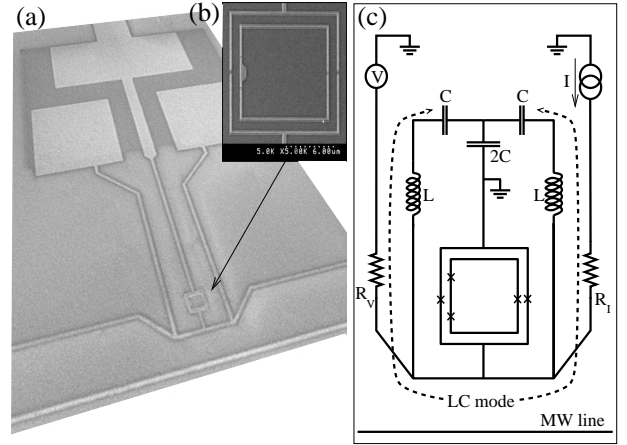}
  \caption{\label{fig1} (a) SEM micrograph of the sample. The qubit and the detector SQUID enclosing it are the small square loops in the lower center picture. The square plates at the top of the picture are the top plates of the on-chip capacitors separated by an insulator from the large bottom plate. (b) A close up of the qubit and the SQUID. The qubit dimension is  $10.2\!\! \times\!\! 10.4\ \mu$m$^2$. (c) Equivalent circuit of the sample. The Josephson junctions are indicated by crosses: three in the inner qubit loop and two in the SQUID. The LC mode is indicated by the dashed line. The inductance and capacitances are calculated from the geometry to be $L=140$~pH and $C = 10$~pF, and the qubit LC oscillator mutual inductance to be $M = 5.7$~pH. The current and voltage lines are filtered through a series combination of copper powder filters and lossy coaxial cables at mixing chamber temperature and on--chip resistors ($R_\mathrm{V}=3$ k$\Omega$ and $R_\mathrm{I}=1$ k$\Omega$).}
\end{figure}

\begin{figure}[bht]
  \includegraphics[width=0.95\linewidth]{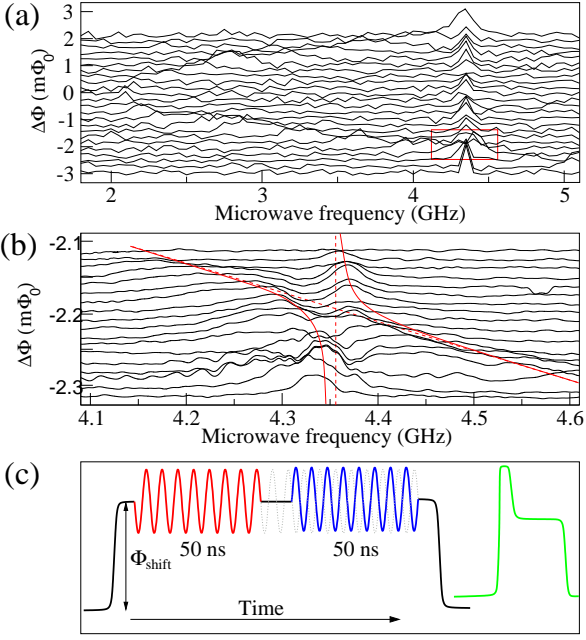}
  \caption{\label{fig2} (a) Spectroscopic characterization of the qubit--oscillator system showing the LC oscillator at $\nu_\mathrm{r}=4.35$~GHz and the qubit dispersion around the gap of $\Delta = 2.1$~GHz.  (b) A close up of the region around 4.35~GHz (indicated by the red square in  (a) showing an avoided crossing. The lines are guides for the eye. (c) Schematic of the pulse sequence used to obtain the spectroscopy in (b): the qubit operating point is fixed at 10.5 GHz and via the MW line a shift pulse of variable height moves the operating point to the vicinity of 4.35 GHz. Two 50 ns MW pulses separated by 2 ns are added to the shift pulse. Here the second MW pulse is phase shifted by 180$^\circ$. The phase shifted second pulse damps the oscillations in the LC circuit~\cite{trams}, and is crucial in terms of resolving the relatively weak qubit signal in this region. After the MW pulses the qubit state is measured by applying a measurement pulse to the SQUID (green curve). The spectroscopy in Fig.~2(a) was obtained with the same scheme, but without the second phase shifted pulse.}
\end{figure}

In this letter we report the observation of vacuum Rabi oscillations in a  macroscopic superconducting solid state system. In our strongly coupled system, the qubit state and the qubit interaction time with the oscillator are controlled by a combination of microwave and DC-shift pulses, resulting in a measuring sequence analogous to atomic cavity QED. We demonstrate vacuum Rabi oscillations between two quantum objects with {\it macroscopically} distinct states\cite{leggett}:  the states are the coherent motion of the superconducting condensate involving millions of Cooper pairs, or the order of a few 100 nA in terms of current associated with the states.

Superconducting Josephson junctions based qubits are ideal artificial atoms where quantum coherence has been demonstrated for many different types\cite{Naka,Vion,Martin,Yu,Chior,duty,kuts}. Recently, two qubit operations in the time domain have also been demonstrated\cite{pashkin,yamamoto,mcdermott}. In our experiment a superconducting persistent current qubit plays the role of the atom. The persistent current qubit is a superconducting loop interrupted by three Josephson junctions (see Fig.~1) with one junction smaller by a factor of $\alpha=0.8$. In the vicinity of half a flux quantum in the loop the device is an effective two--level system\cite{mooij} described by the Hamiltonian $H_{\n{qb}}=h/2\left( \epsilon \sigma_z + \Delta \sigma_x \right)$, given in the basis of clockwise and counterclockwise currents. Here $\sigma_{z/x}$ are the Pauli spin matrices, $h\epsilon=2 I_\n{p} \left( \Phi_{\n{ex}}  -  \Phi_0/2 \right) $ is the energy bias ($I_\n{p}$ is the persistent current in the qubit, $\Phi_{\n{ex}}$ is the external flux threading the qubit loop, and $\Phi_0$ is the flux quantum), and $\Delta$ is the tunnel splitting. The energy gap of the qubit, controlled by the external flux, is $h E=h \sqrt{\epsilon^2+\Delta^2}$. 
\begin{figure}[bht]
  \includegraphics[width=0.95\linewidth]{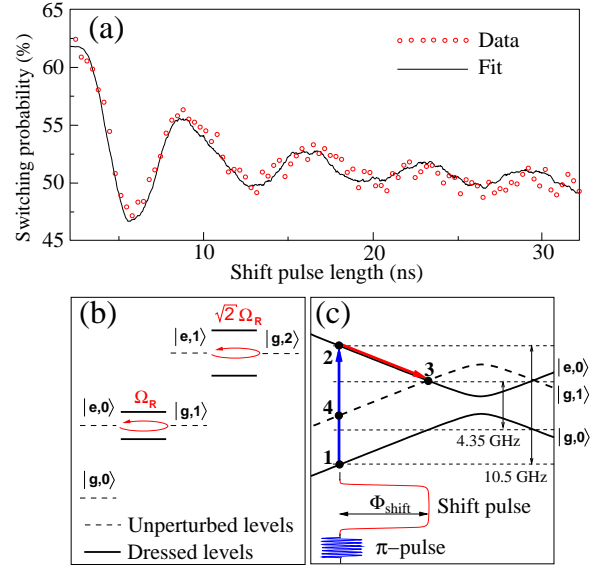}
  \caption{\label{fig3} (a) Vacuum Rabi oscillations (symbols) and a numerical fit (solid line). (b) The few lowest unperturbed and dressed energy levels when the system is in resonance. (c) The qubit energy level diagram and pulse sequence for the vacuum Rabi measurements. The $\pi$ pulse (4.6 ns long at -16 dBm) on the qubit brings the system from state {\bf 1} to  {\bf 2}  and the shift pulse changes the flux in the qubit by $\Phi_\mathrm{shift}$, which, in turn, changes the operating point from {\bf 2} to {\bf 3} where the system undergoes free evolution between $|\n{e},0 \rangle$ and $|\n{g},1 \rangle$ at the vacuum Rabi frequency $\Omega_\mathrm{R}$ until the shift pulse ends and the system returns to the initial operating point where the state is measured to be either in {\bf 2} or {\bf 4}. }
\end{figure}

In our sample geometry, the qubit is spatially separated from the rest of the circuitry. The qubit is enclosed by a superconducting quantum interference device (SQUID) that is inductively coupled to the qubit. The SQUID functions as a detector for the qubit state: the switching current of the SQUID is sensitive to the flux produced by the current in the qubit. The qubit is also enclosed by a larger loop containing on--chip capacitors that provide a well--defined electromagnetic environment for the SQUID and filtering of the measurement leads. The lead inductance $L$ and capacitance $C$ in the outer loop constitute an LC oscillator [see Fig.~\ref{fig1}(b)] with resonance frequency $\omega_\n{r}=2\pi\nu_\n{r}=1/\sqrt{LC}$. The LC oscillator is described by a simple harmonic oscillator Hamiltonian: $H_{\n{osc}} = \hbar \omega_\n{r} \left(a^\dagger a + 1/2 \right)$,  where $a^\dagger$ ($a$) is the plasmon  creation (annihilation) operator. The qubit is coupled to the LC oscillator via the mutual inductance $M$, giving an interaction Hamiltonian $H_\n{I} = h \lambda \sigma_z \left( a^\dagger + a \right)$, where the coupling constant is $h \lambda = M I_\n{p} \sqrt{\hbar \omega_\n{r} /2 L}$. The total system is thus described by a Jaynes-Cummings\cite{cQEDbook} type of Hamiltonian $H = h/2\left( \epsilon \sigma_z + \Delta \sigma_x \right) + \hbar \omega_\n{r} \left(a^\dagger a + 1/2 \right) + h \lambda \sigma_z \left( a^\dagger + a \right)$. We denote  the state of the system by $|Q,i\rangle$, with the qubit either in the ground ($Q=\n{g}$) or excited  ($Q=\n{e}$) state, and the oscillator in the Fock state ($i=0,1,2,\dots$). The parameters of the system can readily be engineered during fabrication; the qubit gap is determined by $\alpha$ and the junction resistance, the oscillator plasma frequency is fully determined by $L$ and $C$, and the coupling between the qubit and the oscillator can be tuned by $M$ and $L$. 

The measurements were performed in a dilution refrigerator ($T\approx20$~mK) using strongly filtered measurement lines and low noise electronics. The system was initially allowed to relax to its ground state, where the thermal occupation of the excited states is negligible ($k_\n{B} T \ll h E, \hbar \omega_\n{r}$). The readout of the qubit state is performed by applying a pulse sequence  to the SQUID [see Fig.~2(c)], and recording whether the SQUID had switched to a finite voltage state or had remained in the zero voltage state. The height and length of the pulse were adjusted to give the best discrimination between the ground and excited states. In our configuration the measured switching probability is directly related to the occupation of the excited state. Coherent control of the system is achieved by applying microwave (MW) to the system via the MW line (see Fig.~1). The MW line is inductively coupled to the system, and coherent MW excitations introduce an oscillating driving term in the Hamiltonian: $H_\n{D} = B_{\n{qb}}  \sigma_z\cos{\omega_\n{ex} t} +  B_\n{osc} \left( a^\dagger + a \right)\cos{\omega_\n{ex} t}$, where $B_\n{qb}$ ($B_\n{osc}$) is the amplitude of the qubit (LC oscillator) drive and $\nu_\n{ex} = \omega_\n{ex}/2\pi$ is the drive frequency. The amplitudes of the drive are proportional to the amplitude of the MW signal, and because of the geometry the LC oscillator experiences a four times larger amplitude, $B_\n{qb} = 1/4\ B_\n{osc}$.

We start with a spectroscopic characterization of the system to map out the qubit and LC oscillator  parameters. The large scale sweep in Fig.~2(a)  shows the qubit dispersion curve and LC oscillator at $\nu_\n{r} = 4.35$~GHz. The qubit parameters  $\Delta=2.1$~GHz and $I_\n{p}=350$~nA are extracted from the dispersion curve. Figure~2(b) shows spectroscopic measurements obtained in the vicinity of the resonant frequency of the LC oscillator (area enclosed by the red square in Fig.~2(a) where an avoided crossing can be seen. Here, the LC oscillator is strongly excited to a classically oscillating state and we cannot draw any quantitative conclusion about the coupling from the spectroscopy, {\it i.e.,} we are not able to observe the vacuum Rabi splitting which usually refers to the splitting of the states involving the two lowest oscillator Fock states\cite{spara}.

Next we investigate the dynamics of the coupled system in the time domain. We performed a measurement cycle where we first excited the qubit and then brought the qubit and the oscillator into resonance where the exchange of a single energy quantum between the qubit and oscillator manifests itself as the vacuum Rabi oscillation $|\n{e},0\rangle \leftrightarrow |\n{g},1\rangle$ (see Fig.~3). Figure.~3(c) is a schematic of the pulse sequence: We started by fixing the qubit operating point far from the resonance point [point 3 in Fig.~\ref{fig3}(c)] and prepared the qubit in the excited state by employing a $\pi$ pulse. The $\pi$ pulse was followed by a shift pulse, which brought the qubit into resonance with the oscillator for the duration of the shift pulse. After the shift pulse the qubit and the oscillator were brought back into off--resonance and the measurement pulse was applied to detect the state of the qubit. It is important to note that the rise time of the shift pulse, $\tau_\n{rise}=0.8$~ns, is adiabatic with respect to both the qubit and the oscillator, $\tau_\n{rise} > [1/E, 2\pi/ \omega_\n{r}]$, but non-adiabatic with respect to the coupling of the two systems, $\tau_\n{rise} < 1/\lambda$. Hence, when the system reaches the resonant point, it is in the state $|\n{e},0\rangle$, which is not an eigenstate of the total Hamiltonian and therefore free evolution between the states $|\n{e},0\rangle$ and $|\n{g},1\rangle$ begins. The physics behind the vacuum Rabi oscillations is thus different from that of normal Rabi oscillations where the system is driven by an external classical field and oscillates between two energy eigenstates. Also, in the normal Rabi oscillations the Rabi frequency is determined by the drive amplitude whereas the vacuum Rabi oscillation frequency is determined only by the system`s intrinsic parameters. The observed Rabi oscillations are in excellent agreement with those calculated numerically [solid line in Fig. 3(a)]. The numerical calculation uses the total Hamiltonian and incorporates the effects of the decoherence of the qubit and the oscillator. The calculation was performed with the known qubit and LC oscillator parameters (obtained from spectroscopy and qubit experiments: qubit dephasing rate $\Gamma_\phi=0.1$~GHz, qubit relaxation rate $\Gamma_e=0.2$~MHz, $\Delta=2.1$~GHz, $\omega_r/2\pi=4.35$~GHz) and by treating the coupling constant and oscillator dephasing and relaxation rates as fitting parameters. From the fit we  extracted the coupling constant $\lambda = 0.2$~GHz, oscillator dephasing rate $\Gamma_\phi=0.3$~GHz,  and relaxation rate $\Gamma_\mathrm{e}=0.02$~GHz. The coupling constant extracted from the fit agrees well with that calculated with the mutual inductance $\lambda=0.216$~GHz.
\begin{figure}[bht]
  \includegraphics[width=0.95\linewidth]{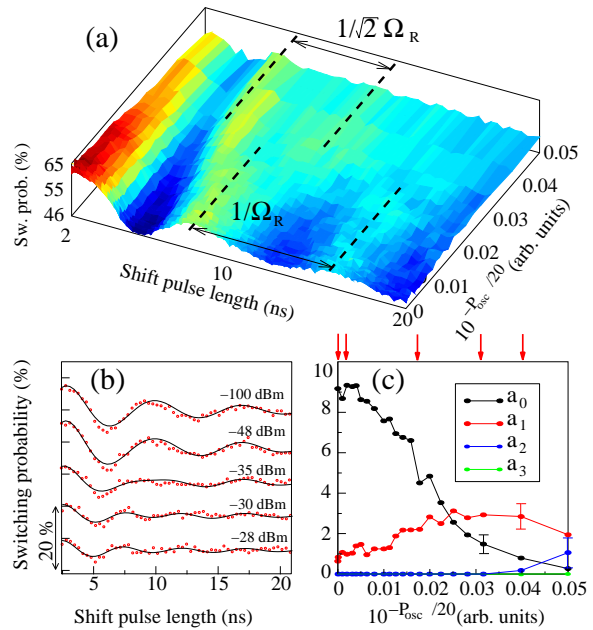}
  \caption{\label{fig4} (a) Rabi oscillations when a 2 ns long pulse with frequency $\nu_\mathrm{ex} =4.35$ GHz and amplitude $A_\mathrm{MW} \propto 10^{P_\mathrm{osc}/20}$ is inserted between the $\pi$ pulse and the shift pulse. (b) Measured Rabi oscillations at different drive powers (symbols), and a fit (solid curve) to $\left[\sum_{n=0}^3 a_n \cos(\sqrt{n+1}\ \Omega_\mathrm{R} t)\right]\exp(-\Gamma t)$ with $a_0,\dots ,a_3$ and $\Omega_\mathrm{R}$ as the only fitting parameters ($\Gamma$  is fixed from a fit to the {-100 dBm} curve). (c) The weights of the different frequency components $a_0,\dots ,a_3$ obtained from the fit as a function of the drive amplitude (The red arrows show the position of the curves in (b). The error-bars indicate the errors obtained from the fitting procedure.}
\end{figure}

To confirm that the oscillations we observed  originated from the coupling of the qubit to a {\it quantized} harmonic oscillator we performed an experiment where we actively excited the oscillator to probe its quantized nature, along the lines of Ref.~\onlinecite{Haroche}. We inserted a short (2 ns) pulse that was resonant with the oscillator (4.35 GHz and variable power $P_\n{osc}$) between the $\pi$ pulse and the shift pulse. Figure~4(a) shows the Rabi oscillations for increasing power to the oscillator. The coherent drive of the oscillator excites higher states according to a Poissonian distribution. The Rabi frequency involving Fock states $n$ and $n+1$ is given by $\sqrt{n+1}\Omega_\n{R}$, hence the Rabi oscillations become faster when the oscillator is in higher excited states, as seen at the top of Fig.~4(a), and there are several frequency components in the oscillations, which can be seen in Fig.~4(a): a crossover of the vacuum Rabi frequency $8:5.5 \approx 1:1/\sqrt{2}$ as the amplitude is increased. In Fig.~4(b) the Rabi oscillations are plotted together with a fit to an exponentially decaying sum of cosines with frequency, $\sqrt{n+1}\Omega_R$ and $n=0,\dots,3$. The only fitting parameters are the amplitudes of the cosines and the Rabi frequency $\Omega_R$. The decay rate being determined from the curve with the lowest power (-100 dBm). The Rabi frequency obtained from the fits varied between 137 and 147 MHz, which is within the accuracy of the fitting. In Fig.~4(c) the amplitudes of the cosines are plotted as a function of the drive amplitude $A_\n{MW} \propto 10^{P_\n{osc}/20}$. When the drive amplitude increases the population of the oscillator ground state decreases (the populations of the states are proportional to the amplitudes of the frequency components)  and the excited states become increasingly populated. We observe significant contributions to the Rabi frequency from the three lowest oscillator states. With a stronger drive to the oscillator the Rabi signal disappears, probably due to faster decoherence from the higher excited states. In the case of a qubit coupled to a classical oscillator there would be  one single frequency component of the Rabi oscillations, which is not the case in our measurements where we can map out up to three frequency components, showing that our system is in the quantum regime in which the discreteness of quantized energy levels is manifest. When the drive to the oscillator is very weak or non--existent the first excited state still has a finite population. We attribute this population to the strong coupling between the MW--line and the oscillator; the oscillator is excited even by off-resonant microwaves. This is confirmed by our numerical calculations, in the  fit of the vacuum Rabi oscillation [Fig.~3(a)] we found $\langle 0 \rangle \approx 0.8$ and $\langle 1 \rangle \approx 0.2$ after the $\pi$ pulse. Therefore we can conclude that the Rabi oscillations observed without excitation of the oscillator are predominantly related to the ground and first excited state of the oscillator, that is, the vacuum Rabi oscillation.  With improved design, the coupling of the oscillator to the MW--line can easily be reduced by more than one order of magnitude. We also note that the relaxation and dephasing times were shorter for the oscillator than for the qubit. This can partly be attributed to the fact that the oscillator was galvanically connected to the measurement lines and we believe that in a free standing LC circuit the relaxation and dephasing times will be much improved.

In conclusion ,we have demonstrated vacuum Rabi oscillations in a superconducting solid--state system, showing the exchange of a single energy quantum between a qubit and a LC circuit. We have also used the qubit to probe the state of the LC oscillator and showed that the LC circuit displayed characteristics of level quantization. With improved sample design it should be possible to generate and measure nonclassical states of the LC oscillator, as have been done in ion trap experiments~\cite{oscStates}. Our results supports the suggestions that future quantum information processing in solid--state devices can be achieved with this kind of system, for example coupling several qubits via an oscillator\cite{makhlin,dansk,ima}.

\begin{acknowledgments}
We thank S. Hughes, H. Kamada, Y. Yamamoto, and I. Chiorescu  for very helpful discussions.
\end{acknowledgments}


\begin{thebibliography}{99}
\bibitem{trams} The oscillations in the LC circuit can be damped  by the out of phase pulse because the first MW pulse produces a coherent classical oscillating state in the LC circuit. The qubit, on the other hand, is in a mixed state (pulse length longer than qubit coherence time) with no phase information after the first pulse, and hence is not affected by the second pulse.
\bibitem{spara} The oscillator being in a classical state of does not contradict the observation of an avoided crossing. It is known that split resonance peaks can be explained by both classical\cite{zhu} and quantum models\cite{cQEDbook}, which sometimes give identical results\cite{steven}.
\end{thebibliography}
\end{document}